\documentclass[12pt]{article}
\usepackage{graphicx}
\usepackage{amssymb,amsmath,amsfonts,palatino,amsthm}
\usepackage{amssymb}
\usepackage{epstopdf}
\newcommand{\f}{\begin{equation}}
\newcommand{\ff}{\end{equation}}

\begin{document}

\title{ Time asymmetric  extensions of general relativity}
\author{
Marina Cort\^{e}s${}^{1,2,3}$, Henrique Gomes${}^{1}$ and 
Lee Smolin${}^{1,4}$ 
\\
\\
Perimeter Institute for Theoretical Physics${}^{1}$\\
31 Caroline Street North, Waterloo, Ontario N2J 2Y5, Canada
\\
\\
Institute for Astronomy, University of Edinburgh ${}^{2}$\\
Blackford Hill, Edinburgh EH9 3HJ, United Kingdom
\\
Centro de Astronomia e Astrof\'{i}sica da Universidade de Lisboa${}^{3}$\\
Faculdade de Ci\^encias, Edif\'{i}cio C8, Campo Grande, 1769-016 Lisboa, Portugal
\\
\\
Department of Physics and Astronomy, University of Waterloo${}^{4}$
\\
Department of Philosophy, University of Toronto${}^{4}$
}
\date{\today}

\maketitle

\begin{abstract}

We describe a class of modified gravity theories that deform general relativity in a way that breaks time reversal invariance and, very mildly, locality. The algebra of constraints, local physical degrees of freedom, and their linearized equations of motion, are unchanged, yet observable effects may be present on cosmological scales, which have implications for the early history of the universe.

This is achieved in the Hamiltonian framework, in a way that requires the constant mean curvature gauge conditions and is, hence, inspired by shape dynamics.

\end{abstract}

\newpage

\tableofcontents

\section{Introduction}

The universe is highly asymmetric in time.  This is especially puzzling because the laws of physics that are up till now well confirmed by experiment are invariant under time reversal.  To reproduce a universe with strong arrows of time appears to require that these time reversal invariant laws are supplemented by highly time asymmetric cosmological initial conditions.  These are needed to ensure that the universe starts off extremely homogeneously and with low entropy.    But this just pushes the question back to the early universe.  What chose those initial conditions? 

Another kind of approach  to the origin of time asymmetry was put forward in 1979 by Penrose\cite{Roger} when he proposed that {\it the fundamental laws are asymmetric under time reversal.  These laws presumably include quantum gravity, which therefor dominates near initial and final singularities.  These may be approximated by time symmetric laws at low energies and away from singularities.   But near singularities the time asymmetry of the fundamental laws manifests itself as highly time asymmetric restrictions on initial conditions which are imposed at initial singularities and not final singularities.} 
We can call this Penrose's hypothesis\footnote{More specifically, Penrose hypothesizes that the Weyl curvature vanishes for initial, but not for final, singularities.  We will here focus on the more general observation that the initial conditions must be time asymmetric.}.

Thus, this paper  develops a temporal physics - a physics in which time is fundamental and not emergent as discussed in \cite{TR,SURT,TN}.   Other approaches to a time asymmetric fundamental physics are developed in \cite{ECS1,ECS2}.

But if general relativity is emergent from an underlying cosmological theory which is time asymmetric, this should manifest itself, within an effective field theory perspective, as a time asymmetric extension of general relativity.  

But there is no known way to break time reversal symmetry within the framework of local, space-time diffeomorphic theories of the space-time metric.  There are no local terms which can be added to the action of general relativity which preserve infinitesimal space-time diffeomorphisms but which break the diffeomorphisms which bring about time reversal.   

Thus to introduce a breaking of time reversal symmetry into general relativity we have to modify or break either locality, space-time diffeomorphism invariance, or both.  But we must do  this in a manner that preserves the diffeomorphism invariance of local equations of motion.  Otherwise we risk introducing local degrees of freedom in addition to the two massless spin two helicity states, a step that may bring us into discord with experiment such as the rate of decay of binary pulsars.   Thus we seek to modify general relativity in a way that preserves the number of degrees of freedom and the general structure of the theory, but yet gives us a means to relate cosmological effects to a breaking of time-reversal symmetry.  This should,

\begin{enumerate}

\item{}not affect the counting of local degrees of freedom or the linearized equations of motion they satisfy.  This means that the algebra of the constraints must be preserved, even if their interpretation as generators of local space-time diffeomorphisms is modified.

\item{}be non-local and cosmological.  After all, quantum gravity is expected to be non-local, so it is reasonable to contemplate controlled departures from locality in effective field theories which aim to capture leading order effects of quantum gravity.  So it makes sense to couple time asymmetry to controlled departures from locality.  

\item{}be dominant in the early universe and perhaps be detectable in the present universe.

\end{enumerate}

One way to do this is the topic of this paper, and is described in the next section.
Then, in section 3 we specialize to $FRW$ universes and, in section 4, describe how the new modified gravity theories can give rise to effects that mimic aspects of dark energy, dark curvature and dark radiation, but with time asymmetries, so a collapsing universe can be distinguished from the time reverse of an expanding universe.

\section{Time asymmetric extensions of general relativity}

We will seek to construct extensions of general relativity which break time reversal invariance in a way that satisfies the conditions just stated.  We will do this by working in the Hamiltonian formulation, in a way which is inspired by the framework of shape dynamics\cite{SD1,SD2,linking}.

\subsection{Background}

We begin with the action for general relativity in Hamiltonian form\cite{ADM},
\f
S^{GR}= \int dt \int_\Sigma \left \{   \pi^{ab} \dot{g}_{ab} - {\cal N} {\cal H}^{ADM} - N^a {\cal D}_a
\right \}
\ff

The Hamiltonian constraint of general relativity is\footnote{We work in units where $c=1$ but 
$G$ and $\hbar$ are kept explicit.}
\f
{\cal H}^{ADM} = -\frac{1}{G}  \sqrt{g} (R - 2 \Lambda ) +\frac{G}{\sqrt{g}} (\pi^{ab} \pi_{ab} -\frac{1}{2} \pi^2) + {\cal H}^{\Psi} =0
\ff
Together with the diffeomorphism constraint
\f
{\cal D}_a = { D}_b \pi_a^{b} + {\cal D}^{\Psi}_a =0
\ff
this forms a first class algebra.
Here $\pi^{ab}$, a tensor density, is the  canonical momenta to the three metric $g_{ab}$ and we will be interested in its trace,
\f
\pi  = g_{ab}\pi^{ab}
\ff

The terms $ {\cal H}^{\Psi} $ and ${\cal D}^{\Psi}_a $ describe matter.  We discuss these below.

To get well defined evolution equations we must fix the gauge by imposing a gauge fixing condition,
$ {\cal F}=0  $.  We will be concerned here with the many fingered time gauge invariance, also called refoliation invariance.  We  don't want to touch the spatial  diffeomorphism invariance so
${\cal F}$ should be invariant under spacial diffeomorphisms.  
A good gauge condition requires each gauge orbit satisfies it once.  This can be done locally if the following transversality condition is satisfied
\f
Det \left ( \{  {\cal H}^{ADM} (x)   ,  {\cal F} (y)  \right ) \neq 0
\ff
Nothing more is required of a gauge fixing term, indeed in general the gauge fixing terms $\cal F$ are second class.  

As our concern is cosmology we work in the spatially compact case.
A good gauge condition in this case is the constant mean curvature condition (CMC), expressed by
\f
{\cal F}= {\cal C}:= \pi - \sqrt{g} <\pi > =0 
\label{S}
\ff
where $<\rho >$ for any density $\rho$ is its spatial average
\f
<\rho > =\frac{\int_\Sigma \rho}{\int_\Sigma \sqrt{g}}
\ff
and $V=\int_\Sigma \sqrt{g}   $ is the spatial volume.
The spatial average of a function $f$ is defined by $<f> = <\sqrt{g} f >$.

As ${\cal C}$ is a gauge fixing of ${\cal H}^{ADM} (N)$ they are second class with respect to  each other.  But it is interesting to note that $\cal C$ and ${\cal D}^a$ also form a first class system\cite{SD2,linking,HG-unique}.   Hence there are two systems of four first class constraints- $({\cal H}, {\cal D}_a)$ and $({\cal C}, {\cal D}_a)$ which partially gauge fix each other.

\subsection{A class of consistent time asymmetric modified gravity theories}

The origin of this work was a question asked by one of us\cite{HG-unique}: are there other pairs of systems of four first class constraints which gauge fix each other?   It was found in \cite{HG-unique} that, restricting consideration to constraints  local in the conjugate fields, $g_{ab}$, and $\pi^{ab}$, the pair $({\cal H}, {\cal D}_a)$ and $({\cal C}, {\cal D}_a)$ are unique, up to the addition of a term added to $\cal H$ linear in $\pi$,
\f
{\cal H}^{HG} = -\frac{1}{G} \sqrt{g} (R - 2 \Lambda ) +\frac{G}{\sqrt{g}} (\pi^{ab} \pi_{ab} -\frac{1}{2} \pi^2)
+\frac{1}{L} \pi     + {\cal H}^{\Psi}   =0
\label{H1}
\ff
The new term in $\pi$ is asymmetric under time reversal, under which 
\begin{eqnarray}
t & \rightarrow  -t  \\
g_{ab}  & \rightarrow  g_{ab}  \\
\pi^{ab} & \rightarrow  -\pi^{ab}
\end{eqnarray}
whereas the rest of the hamiltonian constraint is symmetric under time reversal.

Dimensionally, the metric is dimensionless and $\pi^{ab}$ has dimensions of $ \frac{mass}{l^2}$, as then does $\pi$.  Then $L$
is a length, which marks the scale of time asymmetric effects. 

We can further modify this by making that length a function of other dynamical variables. If time irreversible effects are to be cosmological, we can consider giving up locality.  We can then imagine that this is a function of spatially averaged quantities, for example, the volume, $V$, $<R>$  or $<\rho >$ . 

The first case turns out to be especially simple, hence the subject of this paper is the  set of modified theories defined by
\f
{\cal H}^{mod} = - \frac{1}{G}  \sqrt{g} (R - 2 \Lambda ) +\frac{G}{\sqrt{g}} (\pi^{ab} \pi_{ab} -\frac{1}{2} \pi^2)
+f(V) \pi  + {\cal H}^{\Psi}   =0
\label{HM}
\ff
together with the $CMC$ gauge condition, (\ref{S}).  Here $f(V)$ is some fixed function of the volume of each $CMC$ slice.  

It is straightforward to show that in the presence of the CMC gauge condition (\ref{S}), the algebra of constraints remains first class
\f
\{ {\cal H}^{mod}(N), {\cal H}^{mod} (M) \} = \{ {\cal H}^{ADM}(N), {\cal H}^{ADM} (M) \} +
\frac{3}{2}
ff^\prime V \left ( \pi (N) <M> - \pi (M) <N> 
\right )
\label{HMbracket}
\ff
where 
\f
\pi (N) = \int_\Sigma \pi N
\ff
Here we use,
\f  
\{ {\cal H}^{ADM} (N ),f(V)\}= \frac{G }{2} f'(V)\pi(N ), \ \  \{\pi(N ),f(V)\}=\frac{3}{2} V < N >
\ff

Note that the second term of (\ref{HMbracket}) vanishes in the presence of the CMC constraint, (\ref{S}) which 
implies $\pi (N) = V<N> <\pi >$.

Indeed, even without the CMC gauge fixing, that term vanishes so long as the two Hamiltonian constraints generate 
``pure gauge transformations", corresponding to refoliations in which
\f
<N>=<M> =0
\ff 
Hence we have on either ${\cal  C}=0$ or $<N>=<M> =0$, for the particular modification \eqref{HM},
\f
\{ {\cal H}^{mod}(N), {\cal H}^{mod} (M) \} \approx  \{ {\cal H}^{ADM}(N), {\cal H}^{ADM} (M) \} 
\ff


Since this is now a `gauge-fixed theory' (i.e. it contains second class constraints), we need to supplement the theory with an equation for the gauge-fixed lapse:
\begin{equation}
\label{Global_lapse}
\{\mathcal{H}^{mod}(N),\mathcal{C}(x)\}= \left(\nabla^2-\frac{1}{2}\left(3\sigma^{ab}\sigma_{ab}+\langle\pi\rangle^2/2+R+3f(V)\langle\pi\rangle \right)\right)N(x)-\langle\Delta N\rangle
\end{equation}
where $\sigma^{ab}$ is the traceless momenta and the operator inside the mean, and  
$$2\Delta:=3\sigma^{ab}\sigma_{ab}+\langle\pi\rangle^2/2+R+3f(V)\langle\pi\rangle $$
As long as $\Delta>0$ setting \eqref{Global_lapse} to zero is guaranteed to have 
a  positive solution for the lapse (unique up to a multiplicative constant).

The new  term, $f(V) \pi $ introduces effects which are time asymmetric and non-local, but it does so in a way that is minimal and doesn't change the counting of the local degrees of freedom.  Moreover, we can choose $f$ to have a form whereby it is important in the early universe but not in the present universe.  For example we can take
\f
f(V)= \frac{c}{V^p}
\ff
for some positive power $p$.   After a proper into how the theory retains the correct gravitational degrees of freedom in the next subsection (besides implementing relational principles \cite{Flavio_tutorial}),  we study the cosmological implications for different values of $p$.



\subsection{Linearization and physical degrees of freedom}

The most straightforward way to probe the physical degrees of freedom is to study how perturbations of the gravitational field propagate around a homogeneous background. The unrestricted linearized equations for perturbations $(\delta g_{ab}, \delta\pi^{ab})=(h_{ab},p^{ab})$ are:\footnote{As with the closure of the constraint algebra, which already takes into account variations of the constraints, the lapse-fixing equation \eqref{Global_lapse} is taken at the `base point' $(g_{ab},\pi^{ab})$. } 
\begin{equation}\label{delMom}
\delta {\cal D}_a=\pi^{bc}\left(2\nabla_b(h_{ac})-\nabla_a(h_{bc}) \right)+ \nabla_b(p^{bc})g_{ac}
\end{equation}
\begin{equation}
\delta{\cal C}= (g_{ab}p^{ab}
- \pi^{ab}h_{ab})- \sqrt{g}\int d^3 x (g_{ab}p^{ab}
- \pi^{ab}h_{ab})-\frac{h^{ab}g_{ab}}{2}\langle\pi\rangle
\end{equation}
\begin{equation}
\begin{aligned}\label{delScalar}
\delta {\cal H}^{mod} =& -  \left(  \sqrt g(R^{ab} - \frac{1}{2} g^{ab} (R-2\Lambda)) +\frac{2}{\sqrt{g}}( \pi^{ac} \pi_c{}^b - {\frac{1}{2}} \pi \, \pi^{ab}) 
+\frac{g^{ab}}{2\sqrt{g}}\left(\pi^{cd} \pi_{cd} - \frac{1}{2} \pi^2\right)\right)h_{ab}  \\&+(\nabla_i\nabla_jh^{ab}-\nabla^2(h_{ab}g^{ab}))\sqrt{g}+\frac{p^{ab}}{\sqrt{g}}(\pi g_{ab}-2\pi_{ab})\\ 
&+f(V) (g_{ab}p^{ab}
- \pi^{ab}h_{ab}) - \frac{f'(V)\pi}{2}\int d^3x \sqrt{g}g^{ab}h_{ab}
 \end{aligned}
\end{equation}
The difference from the standard linearized Einstein equations are in the last line of \eqref{delScalar}. In the homogeneous case, the momenta are pure trace, $\pi^{ab}=\frac{\pi}{3}g^{ab}$,  the Ricci curvature is proportional to the metric, and the Ricci scalar is a constant. This means that $\Delta$ in the lapse fixing equation \eqref{Global_lapse} is a spatial constant, and thus the solution is $N=C$, which we  set to $C=1$.

 It is easy to see then that setting the perturbations $(h_{ij},p^{ij})$ to be traceless we get the same linearized equations as GR around this background, and the linearized constraints are completely solved for transverse traceless  $(h_{ij},p^{ij})$.  This shows that the local physical degrees of freedom are the same traceless-transverse tensors as general relativity.

 For more general perturbations around homogeneous backgrounds we get the difference: 
$$\delta {\cal H}_{hom}^{mod}-\delta {\cal H}_{hom}^{ADM}=f(V)\left(p
- \langle\pi\rangle\sqrt g\left( h- \frac{1}{2}\partial_V(\ln{f(V)})\int d^3x \sqrt{g}h\right)\right)$$
where $h=h_{ij}g^{ij}$ and $p=p^{ij}g_{ij}$.

For a general background, since we are interpreting the linear constraint $\cal C$ as a gauge symmetry, imposing ${\cal C}=\delta {\cal C}=0$ we obtain the difference in the linearized equations from ADM to the present modified theory: 
\begin{equation}
\delta {\cal H}^{mod}-\delta {\cal H}^{ADM}= f(V)\sqrt{g}\left(\int d^3 x (g_{ij}p^{ij}- \pi^{ij}h_{ij})-\frac{\langle\pi\rangle}{2}\left(h^{ij}g_{ij} +\partial_V(\ln{f(V)})\int d^3x \sqrt{g}g^{ij}h_{ij}\right)\right)\end{equation} 
The only local term that contributes to the difference is then $-\frac{1}{2}f(V)\langle\pi\rangle h $.  

In the appendix, we give another argument that the physical degrees of freedom are unchanged.


\subsection{Coupling to chiral fermions and the role of torsion}\label{sec:chiral}

It turns out that the new physics driven by the $f(V) \pi $ term affects the propagation of chiral fermions.
We  hence study the dynamics of a chiral fermion, represented by a two component Weyl fermion field,
$\Psi_A$ and its canonical momenta $\Pi^A$.  These are represented by  terms in the constraint:
\f
{\cal H}^{\Psi} =  \Pi^A e^a_i \sigma^{ i \ \ B}_A D_a \Psi_B
\ff
\f
{\cal D}_a^{\Psi} =  \Pi^A D_a \Psi_A
\ff
where $D_a \Psi_A$ is the left handed space time connection which, very importantly, 
{\it depends on $\pi^{ab}$}.
\f
D_a \Psi_A = \partial_a \Psi_A + A_{a A}^{ \  B} \Psi_B
\ff 
where $A_{a A}^{Ash \  B} = A_{a }^{ \  i} \sigma_A^{i \ B}$ is the canonical Ashtekar
connection.  It is defined by\cite{AA}
\f
A^{\ i}_a = \Gamma_a^i +\frac{\imath}{\sqrt{g}} ( \pi_a^i -e_a^i \pi  )
\ff
where  $\Gamma_{a }^{\  i} $ is the Christofel three-connection.  This is defined such that it has the simple commutation relations of the Ashtekar connection:
\begin{eqnarray}
\{ A_a^i (x) , A_b^j (y ) \} & =& 0  
\\
\{ A_a^i (x) , \tilde{E}^b_j (y ) \} & =& \imath \delta^3 (x,y) \delta_a^b \delta_j^i 
\end{eqnarray}
where $\tilde{E}^b_j (y )$ are the densitized triads of the hypersurface. 

Now, normally the Ashtekar connection is equal to the left handed part of the space-time connection,
defined by
\f
\omega_a^i = \Gamma_a^i + \imath K_a^i 
\ff
where $K_a^i$ is the extrinsic curvature defined by
\f
K_{ab}=\frac{1}{2N} (  \dot{g}_{ab}- \nabla_a N_b - \nabla_b N_a )
\ff
where $\nabla_a$ is the covariant derivative associated to the Christoffel three-connection. But now the relation between $\pi^{ab}$ and $K_{ab}$ has been disrupted by the presence of the new term.  From the variation of the action by $\pi^{ab}$ we find
\f
\pi_{ab}= K_{ab}-K g_{ab} -\frac{1}{2} g_{ab} f(V)
\ff
plus a term billnear in the fermions that comes from the variation of the connection dependence of the fermion terms.  This is well understood, so we ignore it in the following.

Consequently the Ashtekar connection which figures in the chiral spinor's dynamics differs from the self-dual connection of the space-time  by a {\it torsion},  $b_a^i$.
\f
A_a^i = \omega_a^i + b_a^i
\ff
where the torsion is
\f
b_a^i = - \frac{\imath}{2} e_a^i f(V)
\ff

Thus there are {\it two} geometries that govern physics at the space-time level.  There is the conventional space-time geometry given by the space-time metric, $g_{\mu \nu}$, which can be reconstructed, up to gauge symmetry, from $g_{ab}, N$ and $N^a$.  It has a metric, compatible, torsion free connection, whose left hand part is $\omega_a^i$.   This governs the propagation of photons.

Then there is the Ashtekar geometry, given by $A_a^i$, which governs the propagation of chiral fermions.  The two geometries differ by  a torsion, $b_a^i$, which contains information about time irreversible, non-local dynamics.  

This means that when $b_a^i$ is signifiant the polarizations of neutrinos and other chiral fermions will propagate differently than those of photons.  Below we see some examples of this phonemena in cosmological models.  

\subsection{Changes of variables}

An obvious question that may be asked is whether a deformation of the dynamics of general relativity may be just general relativity after a change of variables. The generating functional
$$F[g_{ab},\pi^{\prime ab}]=\int d^3 x (g_{ab}\pi^{\prime ab}+2\frac{h(V)}{G}\sqrt{g})$$
 defines the transformation $\pi^{ab}=\frac{\delta F}{\delta g_{ab}}$ and $g^\prime_{ab}=\frac{\delta F}{\delta\pi^{\prime ab}}$. For $h(V)$ satisfying $f(V)=h(V)+h'(V)V$ we get
\begin{eqnarray}
g_{ab} & \rightarrow & g^\prime_{ab} = g_{ab}
\nonumber
\\
\pi^{ab} & \rightarrow & \pi^{\prime ab} =  \pi^{ ab} + \frac{f(V)}{G} \sqrt{g} g^{ab}
\label{cov}
\end{eqnarray}
 That this is a canonical transformation is guaranteed since 
 \f \det\left(
 \begin{array}{rl}
 \frac{\delta g^\prime_{ab}(x)}{\delta g_{cd}(y)} & \frac{\delta g^\prime_{ab}(x)}{\delta\pi^{cd}(y)}\\
  \frac{\delta \pi^{\prime ab}(x)}{\delta g_{cd}(y)} & \frac{\delta \pi^{\prime ab}(x)}{\delta\pi^{cd}(y)}
   \end{array}
\right)=\delta(x,y)\delta^{cd}_{ab} \ff
(since the matrix is block diagonal with vanishing upper right block) which guarantees that the symplectic form stays the unaltered. 

The effect of the transformation is 
\begin{eqnarray}
{\cal H}^{mod} &=&  -\frac{1}{G}  \sqrt{g} (R - 2 \Lambda ) +\frac{G}{\sqrt{g}} (\pi^{ab} \pi_{ab} -\frac{1}{2} \pi^2)
+f(V) \pi   
\nonumber \\
& \rightarrow & - \frac{1}{G}  \sqrt{g} (R - 2 \Lambda^\prime (V)  ) +\frac{G}{\sqrt{g}} (\pi^{ab} \pi_{ab} -\frac{1}{2} \pi^2)
\\
{\cal D}_a   & \rightarrow & {\cal D}_a 
\\
{\cal H}^{\Psi }  & \rightarrow & {\cal H}^{\Psi} - 2 \imath  \alpha f \Pi^A \Psi_A
\\
{\cal D}^{\Psi }_a  & \rightarrow & {\cal D}^{\Psi }_a - 2 \imath  \alpha f \Pi^A e_{a A}^{\ B} \Psi_B
\end{eqnarray}
where the term linear in $\pi$  has been eliminated, but a volume dependence has been added to the cosmological constant (which is no longer constant)
 \f
 \Lambda^\prime (V) = \Lambda + 3  f^2 (V)
 \ff
This gives an effective dark energy which is a function of the global, slicing dependent observable $f$.
Note that the sign of this effective dark energy is necessarily positive.

In addition, as we have just seen,  there is another effect, which is to
add terms to the energy density and pressure imparted by a chiral fermion field.  This is coded in the torsion term, $b_a^i$.  For these two reasons we conclude that the new modified theories are not just GR written in unusual variables.\footnote{Of course, the fact that it could not be just GR was already apparent from the fact that the constraints propagated only on a given gauge-fixing surface.}  

\subsection{A first physical picture}

To get a sense of the physics of the modified theory given by 
(\ref{HM}), note that in the presence of the gauge fixing (\ref{S}) it is weakly equivalent to
\f
{\cal H}^{M} \approx  -\frac{1}{G}  \sqrt{g} (R - 2 \tilde{\Lambda} ) +\frac{G}{\sqrt{g}} (\pi^{ab} \pi_{ab} -\frac{1}{2} \pi^2)
=0
\label{HMapprox }
\ff
with a slice dependent effective dark energy
\f
\tilde{\Lambda }= \Lambda -  \frac{1}{2}f(V) <\pi > =  \Lambda - \frac{c}{2} \frac{<\pi >}{V^p } 
\ff
The effect of the new term in ${\cal H}$ is to effect a time dependent universal expansion
\begin{eqnarray}
\dot{g}_{ab}& = &  \{ g_{ab}(x), f(V) \pi(N) \} = \frac{c}{V^p } g_{ab}(x)
\\
\dot{\pi}^{ab}& = &  \{ \pi^{ab}(x), f(V) \pi(N) \} = -\frac{c}{V^p } \pi^{ab}N(x) - \frac{pc}{2 V^{p-1}} \sqrt{g}g^{ab}(x) \pi(N)
\end{eqnarray}

We will next see how this plays out in FRW models.

\section{Time asymmetry in cosmological models}

\subsection{Reduction to homogeneous, isotropic cosmologies}

To see the consequences of all this we derive the reduction to FRW metrics within the ADM action.
The reduction is defined by
\f
g_{ab}= a^2 (t) q^0_{ab}
\label{FRW1}
\ff
where $q^0_{ab}$ is a non-dynamical metric which is flat or constantly curved.
\f
\tilde{\pi}^{ab} =  \frac{1}{3a} \sqrt{q^0} q^{ab}_0  \pi (t)
\label{FRW2}
\ff
The action reduces to
\f
S=  v_0\int dt \left (
\pi \dot{a} - N {\cal C}
\right )
\label{FRWaction}
\ff
where the fiducial volume of the universe is
\f
v_0 = \int_\Sigma \sqrt{q^0}
\ff

The Hamiltonian constraint, with the homogeneous lapse $N=1$ (solution to equation \eqref{Global_lapse} for the present class of data),  generates time reparametrizations
\f
{\cal C} = \frac{G}{2a}   \pi^2 + G g(a) \pi - a^3 V(a)
\ff
where $g(a)= \frac{a}{G} f(a) $ is a function of $a$.

The standard potential $V$ is
\f
V=  \frac{\Lambda}{6G} - \frac{k}{2G a^2} + \frac{4 \pi  \rho_0 }{3a^3 } 
\ff
whereas the new time non-reversal invariant physics comes from the new $g(a) \pi$ term.  Note that we have inserted a factor of Newton's constant to give $g(a)$ the same dimensions as $\pi$, namely 
a mass per area.

We note that the $CMC$ gauge condition is  satisfied in this  case since the momenta are constant densities.  This means that $a$ and $\pi$ are invariant under volume preserving conformal transformations generated by $\cal C$.

We vary first by $\pi$ to find
\f
\frac{1}{N } \dot{a}= G \left ( \frac{\pi}{a}  + g \right )
\label{aeom}
\ff
This gives us
\f
\pi =  \frac{a^2}{N G }H - a g
\label{pieq}
\ff
in terms of the usual Hubble constant, $H=\frac{\dot{a}}{a}$.

If we vary the action by the lagrange multiplier (or lapse), $N$ we find the 
modified Friedmann equation from ${\cal H}=0$, or 
\f
{\cal H}=  a^3 \left ( \frac{1}{2N^2 G } H^2 - \frac{G g^2}{2a^2} - V 
\right ) =0 
\label{C}
\ff
while varying by $a$ gives an equation for $\dot{\pi}$ 
\f
\frac{1}{N} \dot{\pi} =  \frac{G \pi^2}{2a^2} + 3 a^2 V- a^3 V^\prime -G g^\prime \pi
\label{pieom}
\ff

Combining everything, and fixing the lapse, $N=1$, we find
the modified Friedmann equation
\begin{eqnarray}
\left ( \frac{\dot{a}}{a}  \right )^2 &=& 2GV + \frac{G^2 g^2}{a^2}
\\
&=& \frac{\Lambda}{3} -\frac{k}{a^2 } +\frac{8 \pi G \rho_0 }{3 a^3} + \frac{G^2 g^2}{a^2} 
\label{FRW1}
\end{eqnarray}
and the modified acceleration equation
\begin{eqnarray}
\frac{\ddot{a}}{a} &= &  2GV -aGV^\prime +\frac{G^2 gg^\prime}{a}
\\
&=& \frac{\Lambda}{3} -\frac{4 \pi G \rho_0 }{3 a^3}
 +\frac{G^2 gg^\prime}{a}
 \label{FRW2}
\end{eqnarray}

So the modification gives a new contribution to the energy density,
\f
\rho_{new} =  \frac{g^2}{a^2} \left (   \frac{3 G}{8 \pi }  \right )
\ff
as well as a new pressure,
\f
p_{new}= -\frac{1}{3} \left ( \frac{g^2}{a^2} + 2 \frac{g g^\prime}{a} \right ) \left (   \frac{3 G}{8 \pi  }  \right )
\ff
The equation of state is
\f
w = \frac{p}{\rho} = - \frac{1}{3} \left ( 1 + \frac{2a g^\prime}{g}      \right )
\ff
One can check that a cosmological constant is equivalent to $g=a \sqrt{\frac{\Lambda}{3}}$, which gives $w=-1$.   If we choose $g=ca^p$ we find $w = -\frac{1}{3} ( 1 + 2p)$. 

\subsection{Properties under time reversal}

We note that under the operation of time reversal
\f
dt \rightarrow -dt , \ \ \  a \rightarrow a , \ \ \ \pi  \rightarrow -\pi, \ \ \ \dot{a}  \rightarrow -\dot{a},
\ff
all the equations go to themselves except for the term in $g(a)$ in (\ref{aeom}).  This goes instead to
\f
\frac{1}{N } \dot{a}=G \left (  \frac{\pi}{a}  -  g \right )
\label{notaeom}
\ff
But a solution to this does not solve (\ref{aeom}), even if it solves the modified Friedmann and acceleration equations
(\ref{FRW1}) and (\ref{FRW2}).
So because of this term,  if you stop the system at a time and reverse the velocities, it will not trace back the time reverse of the solution up till then.

So we have to be careful and define the theory by the solutions to the two Hamilton's equations (\ref{aeom}) and (\ref{pieom}), which are first  order in time derivatives, and not by the second order Friedmann and acceleration equations, (\ref{FRW1}) and (\ref{FRW2}).

\section{Simple cases}

By positing different scalings for $g(a)$ we get contributions to the Friedmann equation that scales like different kinds of source terms.  

\subsection{Time asymmetric dark energy, $f=c$, a constant}

The simplest case is $f=c$, a constant, so that $g=a c$.  This corresponds to dark energy because the Friedmann equation is
\f 
\left ( \frac{\dot{a}}{a}  \right )^2 = \frac{8 \pi G}{3} \frac{ \rho_0 }{a^3 } - \frac{k}{a^2} + \frac{\Lambda}{3} + G^2 c^2
\label{FRW4}
\ff
Another way to see that the last, time asymmetric term reproduces dark energy is to note that the 
$\dot{a}$ and $\dot{\pi}$ equations of motion at
$N=1$ , and for vanishing $V$, is now
\f
 \dot{a}= G \left (  \frac{\pi}{a}  + a c  \right )
\label{aeom2}
\ff
\f
 \dot{\pi} =  G \left ( \frac{\pi^2}{2a^2}  - c  \pi \right )
\label{pieom2}
\ff

Notice that these are both solved with $\pi =0$ so that
\f
 \dot{a}=  G c a
\label{aeom3}
\ff
which implies
\f
a(t)= e^{G c t}
\ff 
The three geometry mimics that of the the expanding phase of deSitter.  

The spatial metric geometry of the contracting phase of deSitter  also corresponds to a solution to the equations of motion (\ref{aeom2},\ref{pieom2}), given by
\f
a(t)= e^{- Gc t}  , \ \ \ \ \ \pi (t)= -2c  e^{-2 cG t}
\ff 
This would suggest that the theory has a hidden time reversal symmetry.  But this is not the case as can be seen by the following consideration regarding the role of chiral fermions discussed in section \ref{sec:chiral}.

The Ashtekar connection is proportional to $\pi^i_a$ through the relation that the left handed space-time connection is, in the spatially flat case
\f
A_a^i = \imath \pi_a^i = \frac{\imath}{\sqrt{g}} g_{ab} \pi^{bc} e_c^i = \imath \pi \delta_a^i
\ff
The curvature of the Ashtekar connection is 
\f
F_{ab}^i = \pi^2 \epsilon_{ab}^{\ \ \ i}
\ff
In the conventional deSitter space-time, $\pi = H = \sqrt{\frac{\Lambda}{3}}$.  In this time asymmetric dark energy theory, in the expanding phase $\pi =0$ so the Ashtekar curvature vanishes, while 
in the contracting phase $\pi = -2c  e^{-2 c t}$, so the Ashtekar curvature is 
\f
F_{ab}^i =  4c^2 e^{-4ct} \epsilon_{ab}^{\ \ \ i}
\ff

Thus, when we look to the geometry of the Ashtekar connection, we see that the physics of the contracting  phase is not the time reverse of the expanding phase, and neither of them are identical to the deSitter solution of general relativity.  This is important for the physics, because the energy density of a chiral fermion field has a  term of the form 
$\pi \Pi^A \Psi_A$.   This vanishes for the expanding phase, because $\pi=0$,  while for the contracting phase $\pi$ is nonzero and so the energy density is of the form, $ \pi \Pi^A \Psi_A \approx -2c  e^{-2 c t} \Pi^A \Psi_A $.  

This shows both that the time asymmetric theory is physically different from general relativity and that the theory is time asymmetric, in that the time reversal of the expanding phase of time asymmetric dark energy is not the time reversal of the contracting phase.  Note however that the differences are subtle because the spatial metric is the same as in the expanding and contracting phases of the deSitter solution in general relativity, while the dynamical Ashtekar geometry which governs the propagation of chiral fermions is very different.  This difference cannot be ignored because chiral fermions feel the Ashtekar connection, rather than the metric geometry. 

\subsection{Dark curvature, $f = \frac{G \alpha}{V^{\frac{1}{3}}}$ .}

We note that $g=\alpha$, a constant, behaves like curvature, as it adds a term to the Friedmann equation that scales like $\frac{1}{a^2}$.     Note that real $\alpha$ corresponds to negative curvature.

\subsection{Dark radiation,  $f =\frac{ G \alpha}{V^{\frac{2}{3}}}$}

Another simple case is $g = \frac{G \alpha}{a}$,  where $\alpha$ is a constant, which  behaves like radiation, i.e. $w=\frac{1}{3}.$  As $g(a)= \frac{a}{G} f(a) $, this corresponds to 
\f
f(a) =\frac{ G \alpha}{a^2}, \ \ \ \ \ \ \rightarrow f(V) =\frac{G \alpha}{V^{\frac{2}{3}}}
\ff
so the coefficient of $\pi$ in the modified ${\cal H}$ scales as one over area, i.e. {\it holographicaly}.

This case is described by
\f
\pi =  \frac{a^2}{N G }H - \alpha
\label{pieq2}
\ff
\f
\frac{1}{N} \dot{\pi} = G \frac{\pi^2}{2a^2} + 3 a^2 V- a^3 V^\prime +\frac{G \alpha}{a^2} \pi
\label{pieq2}
\ff
or
\f
\frac{1}{N } \dot{a}= G \left (\frac{\pi + \alpha}{a}   \right )
\label{notaeom2}
\ff
 We note that the universe can start empty, with $V=0$.  We can solve for the empty universe,
 with $N=1$,
 \f
{\cal H} = \frac{G}{2a}   \pi^2 + \frac{G \alpha}{a} \pi =  \frac{G \pi}{a} \left (  \frac{\pi}{2} + \alpha \right ) =  0
\label{CV=0}
\ff
\f
\dot{a}= G \left ( \frac{\pi + \alpha}{a}   \right )
\label{notaeom3}
\ff
\f
\dot{\pi} = G \left ( \frac{\pi^2}{2a^2}+\frac{\alpha}{a^2} \right ) 
\pi
\label{pieom3}
\ff
Note that (\ref{CV=0}) has two solutions.  One  is 
\f
\pi (t)=0,  \ \ \ \dot{a}= \frac{G \alpha}{a} 
\ff
which is solved by,
\f
a(t)= \sqrt{2 G \alpha t},  \ \ \ t>0 \,,
\label{sqrtt}
\ff
according to an radiation in an expanding universe.
Or if we want conformal time, $\tau$  we set $N=a$ to find
\f
\pi (\tau )=0,  \ \ \  a(\tau )= G \alpha t, \ \ \ \tau >0
\ff
The other solution is a contracting solution
\f
\pi (t )=-2 \alpha,  \ \ \  a(t)= \sqrt{- 2 G \alpha t},  \ \ \ t<0
\label{sqrtt}
\ff
or again in conformal time
\f
\pi (\tau )=-2 \alpha,  \ \ \  a(\tau )= - G \alpha t, \ \ \ \tau < 0
\ff
The story is the same as that of asymmetric dark energy:  if you look only at the scale factor $a(t)$,
the contracting phase looks to be the time reverse of the expanding phase, but when one looks at the Ashtekar geometry the two are not time reversals of each other because $\pi$, and hence the Ashtekar curvature vanishes in the expanding phase, but is a constant in the contracting phase.
Indeed the connection has discontinuities at $t=0$.

\section{Conclusions}

In this paper we have presented a class of modified gravity theories which extend general relativity by terms in the field equations which are time asymmetric and depart from locality in a controlled manner.  The theory singles out a particular $3+1$ decomposition which is selected by the constant mean curvature gauge condition and introduces a mild dependence on the volume of these spatial slices.  (Hence, the space-time must be spatially compact.) Nonetheless, the algebra of the constraints is unchanged, and the local physical degrees of freedom are still the spin two degrees of freedom of general relativity.   

However, at a cosmological scale there are new effects, which modify the 
Friedmann and acceleration equations.  At first these can be understood as new forms of dark energy, dark curvature or dark radiation, however a closer look shows the equations of motion on a cosmological scale are no longer time reversal invariant.  One way to see this is to study the propagation of chiral fermions, which can distinguish the influence of the Ashtekar connection and differentiate it from the purely metric geometry.  We see from this example, that the physics of the universe as a whole is more than the dynamics of local degrees of freedom, just scaled up.

In a companion paper~\cite{DR} we study the early universe cosmology that begins with an empty universe whose initial expansion is driven by dark radiation.

\section*{ACKNOWLEDGEMENTS}

We would like to thank Lin-qing Chen for discussions and collaboration on related work~\cite{DR}.   We also thank Jacob Barnett, Laurent Freidel, Roberto Mangabeira Unger, Flavio Mercati and Vasudev Shyam for discussions.

This research was supported in part by Perimeter Institute for Theoretical Physics. Research at Perimeter Institute is supported by the Government of Canada through Industry Canada and by the Province of Ontario through the Ministry of Research and Innovation. This research was also partly supported by grants from NSERC, FQXi and the John Templeton Foundation.  M.C.\ was supported by the EU FP7 grant PIIF-GA-2011-300606 and FCT grant SFRH/BPD/79284/201(Portugal).  

\appendix

\section{Physical Degrees of Freedom}

It is not difficult to extend previous analyses from shape dynamic \cite{SD2,linking} to show the local physical degrees of freedom are unchanged.

The theory proposed here consists of second class constraints ${\cal H}^{mod}$ and  ${\cal C}$ and the diffeomorphism constraint ${\cal D}_i$,
\begin{equation}
\label{total_Hamiltonian_2nd_class}
H_{tot}= {\cal H}^{mod}(N)+{\cal C}(\rho)+{\cal D}(v)
\end{equation}
where ${\cal D}(v) = \int_\Sigma {\cal D}_a v^a$.

 A simple count of the physical degrees of freedom  yields the correct number, as both ${\cal H}^{mod}$ and  ${\cal C}$ are scalars and second class among each other. A more robust way to deal with the construction of the theory however, is to apply the same approach as done in shape dynamics \cite{SD2}, in which case one obtains a theory in the full variables $(g_{ab}, \pi^{ab})$, with a global Hamiltonian generating time evolution and linear first class constraints $\cal C$ and ${\cal D}_a$. This modified shape dynamics theory is a gauge-unfixing of the original, gauge-fixed theory \eqref{total_Hamiltonian_2nd_class}. 

The most straighforward way of performing this construction is through the use of a ``linking theory" \cite{linking}. 
We first use the ``Stueckelberg trick''. For this we adjoin the conformal factor $\phi(x)$ and its canonically conjugate momentum density $\pi_\phi(x)$ to the metric phase space and adjoin the constraints $\pi_\phi(x)\approx 0$ to the total Hamiltonian \eqref{total_Hamiltonian_2nd_class}. We then use 
\begin{equation}\label{equ:VPCTKretschmann}
 F=\int_\Sigma d^3x\left(\Pi^{ab}e^{4\hat\phi}g_{ab}+\Pi_\phi\phi\right)
\end{equation}
as a generating functional of canonical transformations that represent spatial rescalings that do not change the total spatial volume.   Capital letters in equation (\ref{equ:VPCTKretschmann}) denote transformed variables and $\hat \phi$ denotes the volume preserving part of $\phi$, i.e. $\hat \phi=\phi-\frac{1}{6}\ln\langle e^{6\phi}\rangle$. 
The canonical transformation of the constraints $\pi_\phi\approx 0$ is
\begin{equation}
 Q(\rho)=\int_\Sigma d^3x\,\rho\left(\pi_\phi-4(\pi-\langle\pi\rangle\sqrt{|g|})\right).
\end{equation} The theory is then trivially invariant under local rescalings of $(g_{ij},\pi^{ij})$
accompanied by shifts of $\phi$, a symmetry generated by the new constraint $Q(\rho)\approx 0$.

The constraint $\cal C$ does not change under this transformation, while the other constraints transform into:
\begin{equation}\label{equ:TphiConstraints}
 \begin{array}{rcl}
  \left.T{\cal H}(N)\right|_{\pi_\phi=0}&=&\int_\Sigma d^3x\,N\left(-8\Delta \hat\Omega+\left(\frac 1 6 \langle\pi\rangle^2-2\Lambda+f(V)\langle\pi\rangle\right)\hat\Omega^5+R\hat\Omega-\frac{\sigma^a_b\sigma^b_a}{|g|}\hat\Omega^{-7}\right)\\ \\
  T{\cal D}(\xi )&=&\int_\Sigma d^3x\left(\pi^{ab}(\mathcal L_\xi g)_{ab}+\pi_\phi\mathcal L_\xi \phi\right),
 \end{array}
\end{equation}
where $\hat\Omega=e^{\hat{\phi}}$ and where we used the first class constraints $\pi_\phi(x)\approx 4(\pi(x)-\langle{\pi}\rangle\sqrt {|g|(x)})$.

For the `shape dynamics' form of the present theory, we then choose a gauge-fixing of the extended theory,  $\pi_\phi=0$.  
 This gauge condition commutes with $ Q$, 
so local rescalings and spatial diffeomorphisms remain as  symmetries. In other words, the condition $\pi_\phi\equiv 0$ simplifies the diffeomorphism  and rescaling constraints to:
\begin{equation}\label{equ:linearConstraints}
 \begin{array}{rcl}
T{\cal D}(\xi )\rightarrow {\cal D}(\xi )&=&\int_\Sigma d^3x\,\pi^{ab}(\mathcal L_\xi g)_{ab}\\
 Q(\rho)\rightarrow {\cal C}(\rho)&=&\int_\Sigma d^3x\,\rho\left(\pi-\langle\pi\rangle\sqrt{|g|}\right).
 \end{array}
\end{equation}
Thus  the constraints ${\cal C}\approx 0$ and the reduction of $Q\approx 0$ are identified with each other.  

 However, $\pi_\phi=0$ does not commute with 
the scalar constraint ${\cal H}^{mod}$, which must therefore be dealt with.
We will not go into the details here (see \cite{linking}), but the reduction of the remaining scalar constraint gives an equation for the unrestricted conformal factor
$\Omega=e^{\phi}$, which has a unique positive solution, $\phi_o[g_{ij},\sigma^{ij},\tau]$, for $2\Lambda -\frac{3}{8}\tau^2+f(V)\tau>0$, where $\tau=\frac 3 2 \langle \pi\rangle$ is the so called `York time'. The evolution in $\tau$ is then finally generated by the physical Hamiltonian
\begin{equation}\label{equ:Hamiltonian18}
 H^{ sd}_{mod}(\tau)=\int_\Sigma d^3x \sqrt{|g|} e^{6\phi_o[g_{ij},\sigma^{ij},\tau]}.
\end{equation} called the `York volume' Hamiltonian of the (modified) shape dynamics theory, which commutes with the remaining linear constraints. 

We now have a theory  with a phase space parametrized by $(g_{ij},\pi^{ij})$ (or equivalently, by $g_{ij},\sigma^{ij},\tau)$), with a global Hamiltonian generating true evolution in York time, and with 
with linear constraints ${\cal H}_i$ and $\cal C$, generating respectively spatial diffeomorphisms and local spatial rescalings, which are seen as the constraints embodying principles of relationalism \cite{Flavio_tutorial,TR,3R,SURT,TN}. This theory has the same  physical degrees of freedom as ADM general relativity, and it can be said to be the 'gauge-unfixing' of the theory \eqref{total_Hamiltonian_2nd_class}, which means that on-shell it generates evolution equivalent to the one generated by the constraint ${\cal H}^{mod}$ with ${\cal C}=0$ and lapse obeying equation \eqref{Global_lapse}.

\end{document}